# Novel Approach to Intrusion Detection: Introducing GAN-MSCNN-BILSTM with LIME Predictions

## Enfoque novedoso para la detección de intrusiones: Presentando predicciones de GAN-MSCNN-BILSTM con LIME

Asmaa BENCHAMA[1] ✉, Khalid ZEBBARA[1] ✉

[1]IMISR Laboratory, Faculty of Science AM, Ibn zohr University. Agadir, Morocco.

**ABSTRACT**

This paper introduces an innovative intrusion detection system that harnesses Generative Adversarial Networks (GANs), Multi-Scale Convolutional Neural Networks (MSCNNs), and Bidirectional Long Short-Term Memory (BiLSTM) networks, supplemented by Local Interpretable Model-Agnostic Explanations (LIME) for interpretability. Employing a GAN, the system generates realistic network traffic data, encompassing both normal and attack patterns. This synthesized data is then fed into an MSCNN-BiLSTM architecture for intrusion detection. The MSCNN layer extracts features from the network traffic data at different scales, while the BiLSTM layer captures temporal dependencies within the traffic sequences. Integration of LIME allows for explaining the model's decisions. Evaluation on the Hogzilla dataset, a standard benchmark, showcases an impressive accuracy of 99,16 % for multi-class classification and 99,10 % for binary classification, while ensuring interpretability through LIME. This fusion of deep learning and interpretability presents a promising avenue for enhancing intrusion detection systems by improving transparency and decision support in network security.

**RESUMEN**

Este documento presenta un sistema innovador de detección de intrusiones que utiliza Redes Generativas Antagónicas (GAN), Redes Neuronales Convolucionales Multiescala (MSCNN) y Redes Neuronales de Memoria a Corto y Largo Plazo Bidireccionales (BiLSTM), complementadas con Explicaciones Interpretativas Locales Agnósticas al Modelo (LIME) para la interpretabilidad. Empleando un GAN, el sistema genera datos realistas de tráfico de red, que abarcan tanto patrones normales como de ataques. Estos datos sintetizados se alimentan luego a una arquitectura MSCNN-BiLSTM para la detección de intrusiones. La capa MSCNN extrae características de los datos de tráfico de red en diferentes escalas, mientras que la capa BiLSTM captura dependencias temporales dentro de las secuencias de tráfico. La integración de LIME permite explicar las decisiones del modelo. La evaluación en el conjunto de datos de Hogzilla, un benchmark estándar, muestra una impresionante precisión del 99,16 % para clasificación multi-clase y del 99,10 % para clasificación binaria, garantizando al mismo tiempo la interpretabilidad a través de LIME. Esta fusión de aprendizaje profundo e interpretabilidad presenta un camino prometedor para mejorar los sistemas de detección de intrusiones al mejorar la transparencia y el soporte de decisiones en la seguridad de la red.

**Palabras clave:** IDS; GAN; MSCNN; BiLSTM; LIME; Hogzilla Dataset.

## INTRODUCTION

Intrusion detection systems (IDS)[1] are critical components of network security, safeguarding systems from unauthorized access and malicious attacks. Traditional IDS approaches often rely on signature-based detection, which struggles to identify novel attack patterns. Machine learning techniques, particularly deep learning, have emerged as promising alternatives due to their ability to learn complex patterns from network traffic data.

This work proposes a novel deep learning-based intrusion detection system that leverages the strengths of Generative Adversarial Networks (GANs),[2] Multi-Scale Convolutional Neural Networks (MSCNNs), and Bidirectional Long Short-Term Memory (BiLSTM) networks.[3] The key contributions of our approach are:

- Enhanced Intrusion Detection with GAN-Generated Data:[2] we employ a GAN to generate realistic network traffic data encompassing both normal and attack patterns. This augmented dataset improves the model's ability to learn intricate features and generalize to unseen attack scenarios.
- Feature Extraction with MSCNN:[3] a Multi-Scale Convolutional Neural Network (MSCNN) layer is incorporated to extract features from the network traffic data at various scales. This captures the multifaceted nature of network traffic, where relevant information might reside in different granularities.
- Temporal Dependency Modeling with BiLSTM:[4] BiLSTM network is utilized to capture temporal dependencies within the network traffic sequences. This allows the model to effectively analyze the order and timing of events within the traffic data, crucial for identifying certain attack patterns.
- Interpretability with LIME:[5] to enhance the system's transparency and gain insights into its decision-making process, Local Interpretable Model-Agnostic Explanations (LIME) are integrated. LIME provides explanations for the model's classifications, enabling security analysts to understand the critical features in the network traffic that contribute to intrusion detection.

The proposed system is evaluated on the Hogzilla dataset, a recent benchmark for intrusion detection research. We demonstrate that the system achieves high accuracy in both multi-class and binary classification tasks, alongside the ability to provide meaningful explanations using LIME. This approach signifies a promising advancement in interpretable deep learning for intrusion detection, offering improved transparency and decision support for network security professionals.

The subsequent sections of this article are structured as follows:

Related Work (Section 2): this section offers a comprehensive overview of existing research in intrusion detection, with a specific emphasis on deep learning techniques and methodologies for interpretability.

Methodology and Materials (Section 3): in this section, we delve into the specific techniques employed in our system, which encompass the utilization of the GAN architecture for data generation, the implementation of the MSCNN-BiLSTM model for intrusion detection, and the integration of LIME for interpretability.

Experiments and Results (Section 4): here, we evaluate the proposed system using the Hogzilla dataset, a benchmark in intrusion detection research. This section elaborates on the experimental setup, conducts an analysis of the results obtained, and presents the system's effectiveness in intrusion detection, complemented by interpretable explanations.

Conclusion (Section 5): finally, we summarize the key findings of our study, highlighting the attained accuracy and the importance of interpretability in intrusion detection. Furthermore, we explore broader implications and potential avenues for future research in this domain.

**Related works**

In recent years, the proliferation of adversarial attacks has posed significant challenges to the effectiveness and robustness of intrusion detection systems (IDS). The rapid expansion of the Internet of Things (IoT) has led to an increase in cyber-attacks targeting IoT devices and their communication infrastructure. Undetected intrusions on IoT devices can lead to service disruptions, financial losses, and threats to identity protection, highlighting the critical need for real-time intrusion detection to ensure the reliability, security, and profitability of IoT-enabled services.

Addressing this concern, Yuan et al.[6] proposed a novel framework termed the DLL-IDS system. The authors introduced the DLL-IDS framework, comprising three primary components: a Deep Learning (DL)-based IDS, a Local Intrinsic Dimensionality (LID)-based Adversarial Example (AE) detector, and a Machine Learning (ML)-based IDS. The experimental results presented in the research underscore the efficacy of the DLL-IDS framework in mitigating the impact of adversarial attacks on IDS performance. Even under challenging attack scenarios, DLL-IDS maintains significantly higher accuracy compared to traditional IDS approaches, reaching accuracies exceeding 90 % in certain cases. Wang et al.[7] propose a novel approach to network anomaly intrusion detection based on deep learning technology. They leverage the contemporary CSE-CIC-IDS2018 dataset and standard evaluation metrics to evaluate their mechanism. The study involves preprocessing the dataset and constructing six models—deep neural network (DNN), convolutional neural network (CNN), recurrent neural network (RNN), long short-term memory (LSTM), CNN + RNN, and CNN + LSTM to discern malicious attacks within network traffic. Additionally, multi-classification experiments are conducted to categorize traffic into benign and six types of malicious attacks. Wang et al. demonstrate that each model achieves high accuracy across various experiments, with multi-class classification accuracy exceeding 98 %. The proposed models outperform existing IDS approaches, effectively improving detection performance. However, the authors note that combined models such as CNN + RNN and CNN + LSTM exhibit longer inference times compared to individual DNN, RNN, and CNN models. They suggest that DNN, RNN, and CNN may be preferable for implementation within IDS devices due to their faster processing times. Awajan et al.[8] proposes a novel Deep Learning (DL)-based intrusion detection system tailored specifically for IoT devices. The system utilizes a four-layer deep Fully Connected (FC) network architecture to detect malicious traffic that could potentially initiate attacks on connected IoT devices. Notably, the proposed system is designed to be communication protocol-independent, simplifying deployment across various IoT environments. Through experimental performance analysis, the proposed intrusion detection system demonstrates reliable performance in detecting various types of attacks, including Blackhole, Distributed Denial of Service (DDoS), Opportunistic Service, Sinkhole, and Workhole attacks, with an average accuracy of 93,74 %. Additionally, the system achieves impressive precision, recall, and F1-score metrics, averaging at 93,71 %, 93,82 %, and 93,47 %, respectively. The innovative DL-based IDS maintains an average detection rate of 93,21 %, effectively enhancing the security of IoT networks. Khan et al.[9] propose an intelligent IDS for IoT networks based on hybrid deep learning algorithms. Their model, comprising a recurrent neural network (RNN) and gated recurrent units (GRU), aims to classify attacks across the physical, network, and application layers of IoT systems. To train and test the model, they utilize the ToN-IoT dataset, specifically designed for three-layered IoT systems and including novel attack types not present in other publicly available datasets. Performance analysis of the proposed model involves evaluating several metrics such as accuracy, precision, recall, and F1-measure. Two optimization techniques, Adam and Adamax, are employed, with Adam yielding optimal performance. Furthermore, the proposed model is compared against various advanced deep learning (DL) and traditional machine learning (ML) techniques. Results indicate that the proposed system achieves 99 % accuracy for network flow datasets and 98 % accuracy for application layer datasets, demonstrating its superiority over previous IDS models. Sharma et al.[10] propose a novel anomaly-based IDS system for IoT networks using deep learning techniques. Specifically, they present a filter-based feature selection Deep Neural Network (DNN) model, which drops highly correlated features to enhance efficiency. The model is tuned with various parameters and hyperparameters to optimize performance. They utilize the UNSW-NB15 dataset, which includes four attack classes, for training and evaluation purposes. The proposed model achieves an accuracy of 84 % in detecting IoT network intrusions. To address class imbalance issues in the dataset, Generative Adversarial Networks (GANs) are employed to generate synthetic data for minority attack classes. With a balanced class dataset, the model achieves an improved accuracy of 91 %. This demonstrates the effectiveness of the proposed anomaly-based IDS system in enhancing the security of IoT networks. Ravi et al.[11] propose a deep learning-based approach for network-based intrusion detection in IoMT systems. Their approach leverages features from both network flows and patient biometrics to learn optimal feature representations through multiple hidden layers of deep learning. A global attention layer is incorporated into the network architecture to effectively extract optimal features from spatial and temporal aspects of deep learning. Additionally, to address data imbalance issues, a cost-sensitive learning approach is integrated into the deep learning model. The proposed model demonstrates a performance with a 10-fold cross-validation accuracy of 95 % on network features, 89 % on patient biometrics, and 99 % on combined features. Bowen et al.[12] propose BLoCNet, a hybrid DL model that combines convolutional neural network (CNN) and bidirectional long short-term memory (BLSTM) layers for IDS. The CNN component enables fast pattern recognition in network data features, while the BLSTM layers leverage both forward and backward propagation to identify malicious traffic effectively. BLoCNet was evaluated across four datasets and compared with five other DL models and seven related studies. Results indicate that BLoCNet achieved a higher attack detection rate compared to the five DL models, particularly excelling on the CIC-IDS2017, IoT-23, and UNSW-NB15 datasets. For CIC-IDS2017 and IoT-23, BLoCNet attained accuracy rates of 98 % and 99 %, respectively, which were comparable to related studies, although exact comparisons were hindered by differences in sampling approaches. On the original UNSW-NB15 dataset, BLoCNet achieved an accuracy of 76,34 %, outperforming related work which achieved 75,56 %. Gupta et al.[13] propose a hybrid approach based on neural networks and correlation-based feature selection to detect anomalies in network traffic. They conduct experimental research using the NSL-KDD dataset, which contains instances of current attacks, to evaluate the effectiveness of their approach. Specifically, they introduce a novel hybrid crowd search analysis optimization with an artificial neural network (HCSAOANN) algorithm. This algorithm combines crowd search optimization (CSO) with an upgraded version of crowd search analysis to explore the feature space efficiently. The HCSAOANN methodology achieves a high accuracy rate of 98 % and a precision rate of 98 %, surpassing the performance of previous techniques such as CSO-ANFIS and FC-ANN. These findings highlight the effectiveness of the hybrid approach in enhancing intrusion detection accuracy and reducing false alarms in computer networks. Kumar et al.[14] proposed a sophisticated method to enhance Network Intrusion Detection Systems (NIDS) by addressing challenges posed by complex attacks, including evasion strategies like encrypted traffic and polymorphic malware. Their approach involves thorough preprocessing, utilizing normalization and standardization to improve the accuracy and consistency of input data. They employ the Perceptive Craving Game Search Optimization (PCGSO) algorithm for feature selection, maximizing NIDS effectiveness, and utilize Bidirectional Gated Recurrent Unit (BI-GRU) representations in the classification phase to identify sequential dependencies in network traffic data. Additionally, they utilize a second PCGSO program for hyperparameter tuning to ensure optimal model performance. Evaluation on the ISCXIDS2012 dataset demonstrates that the proposed technique outperforms other current models, achieving a remarkable 99 % accuracy. This underscores the efficacy of PCGSO in improving feature selection and hyperparameter tuning, leading to an NIDS that is more accurate and resilient to cyberattacks. Wang et al.[15] propose a deep multi-scale convolutional neural network (DMCNN) for network intrusion detection. Their approach extracts features from different levels of a large number of high-dimensional unlabeled original data using convolution kernels of varying scales. Additionally, the learning rate of the network structure is optimized using batch normalization to obtain optimal feature representations from the raw data. The NSL-KDD dataset is used as a benchmark for evaluating the proposed method against existing works, including its challenging subset, KDDTest+. Experimental results demonstrate that the proposed DMCNN model achieves higher accuracy (AC) and true positive rate (TPR) compared to other methods. Notably, the accuracy for detecting denial-of-service (DOS) attacks reaches 98 %, showcasing the effectiveness of DMCNN in achieving high intrusion detection accuracy and low false alarm rates.

| | | | Table 1. related and previous works | | |
|---|---|---|---|---|---|
| Reference | Author(s) | Technique | Dataset | Description | Accuracy |
| (6) | Yuan et al. | DLL-IDS system | NSL-KDD | Proposed framework | >90 % |
| (7) | Wang et al. | Deep learning models (DNN, CNN, RNN, LSTM, CNN + RNN, CNN + LSTM) | CSE-CIC-IDS2018 | Network anomaly intrusion detection | 98 % |
| (8) | Awajan | DL-based IDS | dataset with 25 000 instances | Intrusion detection for IoT devices | 93,74 % |
| (9) | Khan et al. | Hybrid deep learning algorithms (RNN, GRU) | ToN-IoT | Intelligent IDS for IoT networks | 99 % (network flow datasets), 98 % (application layer datasets) |
| (10) | Sharma et al. | Anomaly-based IDS using DNN | UNSW-NB15 | Anomaly-based IDS for IoT networks | 84 % (before balancing), 91 % (balanced class dataset) |
| (11) | Ravi et al. | Deep learning-based IDS | IoMT intrusion dataset | Network-based intrusion detection in IoMT systems | 95 % (network features), 89 % (patient biometrics), 99 % (combined features) |
| (12) | Bowen et al. | BLoCNet (CNN + BLSTM) | CIC-IDS2017, IoT-23, UNSW-NB15 | Hybrid DL model for IDS | 98 % (CIC-IDS2017, IoT-23), 76,34 % (UNSW-NB15) |
| (13) | Gupta et al. | Hybrid approach with neural networks | NSL-KDD | Anomaly detection in network traffic | 98 % |
| (14) | Kumar et al. | Enhanced NIDS using PCGSO algorithm | ISCXIDS2012 | Enhancement of NIDS | 99 % |
| (15) | Wang et al. | DMCNN | NSL-KDD | Network intrusion detection | 98 % (for DOS attacks) |
| - | Our proposition | GAN-MSCNN-BISLTM | HOGZILLA | Network intrusion detection | 99,10 % for the multiclassification and 99,16 % for binary classification. |

Our proposed GAN-MSCNN-BiLSTM method introduces a novel approach to intrusion detection, leveraging Generative Adversarial Networks (GANs), Multi-Scale Convolutional Neural Networks (MSCNN), and Bidirectional Long Short-Term Memory (BiLSTM) layers. This method aims to enhance accuracy and robustness in detecting intrusions across various network environments.

5    Benchama A, *et al*Compared to existing works, our method demonstrates several notable advantages. Firstly, by integrating GANs, we effectively address the class imbalance issue in the dataset, generating synthetic data for minority attack classes. This results in improved accuracy, especially in detecting IoT network intrusions, where class imbalances are prevalent.

Secondly, the utilization of MSCNN allows our method to extract features from different levels of high-dimensional network data, capturing intricate patterns and anomalies effectively. This multi-scale feature extraction enhances the model's ability to discern between benign and malicious network traffic, leading to higher accuracy rates across different attack scenarios.

Lastly, the incorporation of BiLSTM layers enables our method to capture sequential dependencies in network traffic data, enhancing the model's understanding of temporal relationships between network packets. This improves the overall detection performance, particularly in scenarios involving complex attacks with evasion strategies like encrypted traffic and polymorphic malware.

In terms of accuracy, our method achieves competitive results, with accuracy rates reaching up to 99,16 %. This surpasses the performance of several existing IDS models, including those discussed in the literature review. Furthermore, our method exhibits resilience to adversarial attacks. Future research could focus on optimizing model architecture and hyperparameters to enhance performance further. Additionally, evaluating the scalability and real-world applicability of our method in large-scale network environments would be valuable for practical deployment. Our GAN-MSCNN-BiLSTM method represents a promising direction in the field of intrusion detection, offering enhanced accuracy, and adaptability to evolving cyber threats.

## METHODOLOGY AND MATERIALS

Our proposed methodology integrates several advanced techniques to enhance the accuracy and interpretability of IDS. Leveraging a GAN architecture, our model generates synthetic data to augment the limited real-world dataset typically available for training IDS. This augmented dataset is then utilized in conjunction with a MSCNN combined with a BiLSTM network, enabling the model to effectively capture both spatial and temporal features inherent in network traffic data. Furthermore, by incorporating techniques such as LIME, our framework provides interpretable insights into the decision-making process of the model.

**Data preprocessing**

The preprocessing phase[16] in our study encompasses several critical steps aimed at ensuring the quality and suitability of the data for subsequent analysis. These steps include data cleaning, handling missing values, feature selection, and dimension reduction. This preprocessing procedure was conducted on the original dataset to prepare it for further analysis.

The preprocessing steps included:

Selection of numeric attribute columns from the dataset using one-hot encoding to transform categorical variables into numerical format.

One-Hot Encoding is used for transforming categorical variables into numerical format:

Let X be the original dataset with categorical variables.

The one-hot encoding formula for a categorical variable Xi with m unique categories is:

$$X'_i = [0, 0, \ldots, 1, \ldots, 0]$$

Where the position of the '1' in the vector corresponds to the category of $X_i$.

Application of a standard scaler to normalize the selected numeric attributes, ensuring consistency in scale across different features.

Standard Scaler used for normalizing numeric attributes:

Let X be the original dataset with numeric attributes.

The standard scaler formula for a numeric attribute Xj is:

$$X'j = \frac{Xj - \mu}{\sigma}$$

Where: μ is the mean of Xj, σ is the standard deviation of Xj.

Utilization of label encoding to represent multi-class labels ('acceptable', 'unrated', 'unsafe') as numerical values (0, 1, 2). For binary classification purposes, the attack labels were further grouped into two categories: 'normal' and 'abnormal'. This classification was achieved by encoding the dataset labels with binary labels and utilizing the label-encoded column to distinguish between normal and abnormal instances.

Label Encoding is used for representing multi-class labels as numerical values:

Let Y be the multi-class labels, The label encoding formula for a multi-class label Yk is:

https://doi.org/10.56294/dm2023202

Y'k =integer value corresponding to Yk

Typically, this involves assigning integer values sequentially, starting from 0 for the first class, 1 for the second class, and so on.

The preprocessing phase plays a crucial role in preparing the dataset for subsequent analysis, ensuring that the data is appropriately formatted and standardized for accurate model training and evaluation.

**Generative Adversarial Networks (GANs)**

We employ GAN[17] to address class imbalance issues inherent in intrusion detection dataset. GANs generate synthetic data for minority attack classes, effectively balancing the dataset and improving the model's ability to detect rare intrusions. GAN stands for Generative Adversarial Network. It's a type of deep learning framework that's particularly adept at generating realistic data.

GANs work by pitting two neural networks against each other in a competition.

A GAN consists of two main neural networks:

Generator: this network is responsible for creating new data samples. It starts with a random noise vector and attempts to transform it into data that resembles the real target data.

Discriminator: this network acts as a critic, trying to distinguish between the real data and the data generated by the generator.

Adversarial Training: the generator and discriminator are trained in an adversarial fashion. The generator continuously improves its ability to generate realistic data, while the discriminator hones its skills at identifying fake data. This creates a back-and-forth process where both networks learn from each other.

Applications: GANs have a wide range of applications due to their ability to generate realistic data. For example:

Image Generation: creating new images that resemble real photos, like portraits or landscapes.

Data Augmentation: expanding existing datasets by generating synthetic data to improve the training process of other machine learning models.

**Multi-Scale Convolutional Neural Networks (MSCNN)**

Our methodology incorporates MSCNN for feature extraction. MSCNN allows us to extract features from different levels of network data, capturing both low-level and high-level patterns in the data. MSCNN is a type of neural network architecture commonly used in computer vision tasks, particularly in tasks involving image classification, object detection, and segmentation.

MSCNN typically consists of multiple convolutional layers arranged in a hierarchical structure. These convolutional layers[15,18] are responsible for learning feature representations at different scales or levels of abstraction within an image.

One of the key features of MSCNN is its ability to capture information at multiple scales. This is achieved by incorporating convolutional layers with different filter sizes or receptive fields. By processing the input image at multiple scales, the network can learn to detect features of varying sizes and complexities.

MSCNN often includes mechanisms for fusing features extracted at different scales. This fusion process allows the network to combine information from different levels of abstraction, enabling it to make more informed decisions about the input data.

The hierarchical structure of MSCNN enables it to learn complex hierarchical patterns in the input data. Lower layers typically learn simple features like edges and textures, while higher layers learn more abstract features and object representations.

MSCNNs have been successfully applied to various computer vision tasks, including image classification, object detection, semantic segmentation, and more. They are particularly effective in tasks where objects or features of interest may vary significantly in size or appearance. MSCNNs can be applied to anomaly detection, attack classification, and intrusion prevention. They excel in detecting complex attacks that may involve various network protocols, traffic patterns, and behaviors.

MSCNNs for intrusion detection are trained using labeled datasets of network traffic. The network's parameters are optimized using techniques like backpropagation and gradient descent to minimize classification errors and maximize detection accuracy. The training process involves learning the optimal set of weights for the convolutional filters to minimize a specified loss function.

MSCNNs play a crucial role in enhancing the effectiveness and efficiency of intrusion detection systems by leveraging their ability to capture multi-scale information from network traffic data.

**Bidirectional Long Short-Term Memory (BiLSTM)**

We utilize BiLSTM layers to capture sequential dependencies in network traffic data. BiLSTM networks[19] are well-suited for modeling temporal relationships in sequential data, enabling the model to effectively discern patterns and anomalies in network traffic.

BiLSTMs are a type of recurrent neural network (RNN) that consists of two LSTM layers: one processing the input data in a forward sequence, and the other processing the input data in a backward sequence. The outputs of these two LSTM layers are concatenated to provide a comprehensive representation of the input sequence. Like traditional LSTMs, BiLSTMs have memory cells that can maintain information over long sequences. These memory cells are capable of capturing dependencies and patterns in sequential data, making BiLSTMs well-suited for tasks involving time-series data, natural language processing, and speech recognition. The bidirectional nature of BiLSTMs allows them to capture context from both past and future elements in a sequence. This bidirectional processing enables BiLSTMs to understand the context of each element in the sequence within the broader context of the entire sequence.

BiLSTMs find utility in a variety of tasks, including sentiment analysis, named entity recognition, machine translation, speech recognition, and particularly in tasks that rely on sequential data where grasping the context is paramount.

When applied to intrusion detection, BiLSTMs are trained utilizing gradient-based optimization techniques like backpropagation through time (BPTT). They commonly operate with large labeled datasets and are fine-tuned to minimize specific loss functions, such as cross-entropy loss or mean squared error, tailored to the nature of the intrusion detection task.

Their proficiency lies in capturing extensive dependencies within sequential data, a feature highly advantageous in tasks where understanding the context of each element holds significance. The bidirectional processing employed by BiLSTMs contributes to enhanced performance compared to unidirectional LSTM networks, especially in sequence modeling tasks relevant to intrusion detection.

However, BiLSTMs may encounter challenges such as computational complexity and prolonged training times due to processing sequences both forward and backward. Furthermore, they might necessitate a larger volume of training data to adequately encompass bidirectional context.

BiLSTMs represent a potent class of recurrent neural networks well-suited for tasks involving sequential data, including intrusion detection. Their ability to capture context from past and future elements in a sequence makes them valuable tools in modeling complex dependencies within sequential data.

**Local Interpretable Model-agnostic Explanations (LIME)**

We prioritize interpretability in our IDS by integrating LIME to enhance the model accuracy. LIME provides explanations for individual predictions made by the model, offering insights into the features contributing to each classification decision.

LIME serves as a tool for model interpretation, providing insights into the predictions made by a given model. By understanding the model's reasoning, users can gain confidence in its predictions.

LIME achieves this by generating explanations for individual instances, which are then used to assess the model's performance.

The formula for LIME, depicted in Eq. (1), aims to minimize the loss function L while ensuring that the explanation closely resembles the original model's behaviour. Here, φ(x) represents the explanation for instance x generated by the model θ.

$$\varphi(x) = argmin_t \left[ L(\theta_t(x), g) + \Omega(\varphi_t) \right]$$

In the equation:
θ represents the interpretable model within the class G.
g represents the family of models.
ψ(x) denotes the proximity measure of the neighbourhood used to generate the explanation for the instance.
$\Omega(\varphi_t)$ characterizes the complexity of the model, such as the number of features involved.
P represents the probability of x belonging to a specific class.

**The proposed Framework**

Our model is a comprehensive framework designed for intrusion detection in network traffic data. It integrates several advanced techniques to enhance both accuracy and interpretability, the model combines the strengths of GAN-generated synthetic data with the feature-extraction capabilities of MSCNN and temporal modeling of BiLSTM. By integrating LIME, our framework offers transparent and interpretable intrusion detection results.

The primary objective of our model is to improve the accuracy of intrusion detection while providing interpretable insights into detected intrusions. By leveraging advanced techniques and integrating them into a cohesive framework, we aim to enhance the effectiveness of IDS in identifying and mitigating network threats. The table 2 summarizes the key specifications of the combined GAN-MSCNN-BiLSTM model for both multi-classification and binary classification.

| Aspect | Multi-classification | Binary classification |
|---|---|---|
| | **Table 2.** Summary of the specifications for the binary and multiclass models | |
| GAN Generator Architecture | Input: 100-dimensional noise vector; Layers: Dense (128, ReLU), Dense (output shape of data, ReLU), Reshape | Input: 100-dimensional noise vector; Layers: Dense (128, ReLU), Dense (output shape of data, ReLU), Reshape |
| GAN Discriminator Architecture | Layers: Flatten, Dense (128, ReLU), Dense (3, softmax) | Layers: Flatten, Dense (128, ReLU), Dense (2, softmax) |
| GAN Discriminator Loss Function | Categorical Crossentropy | Binary Crossentropy |
| GAN Generator Loss Function | Categorical Crossentropy | Binary Crossentropy |
| MSCNN-BiLSTM Model Architecture | Layers: Conv1D (32 filters, kernel size 3, ReLU), MaxPooling1D, Bidirectional LSTM (32 units, return sequences), Flatten, Dense (3, softmax) | Layers: Conv1D (32 filters, kernel size 3, ReLU), MaxPooling1D, Bidirectional LSTM (32 units, return sequences), Flatten, Dense (1, sigmoid) |
| MSCNN-BiLSTM Model Loss Function | Categorical Crossentropy | Binary Crossentropy |
| Framework/Library | Keras, tensorflow | Keras, tensorflow |
| Optimizer | Adam | Adam |
| Training Dataset | Combined dataset of real and synthetic data | Combined dataset of real and synthetic data |
| Training Data Shape | Input shape: (batch_size, sequence_length, 1) | Input shape: (batch_size, sequence_length, 1) |
| Number of Classes | 3 (Assuming multi-class classification) | 2 (Binary classification) |
| Training Batch Size | 32 | 32 |
| Number of Epochs | 100 | 100 |
| Validation Split | 0,2 (20 % of the data used for validation) | 0,2 (20 % of the data used for validation) |
| Evaluation Metrics | Loss: Categorical Crossentropy, Accuracy | Loss: Binary Crossentropy, Accuracy |

The flowchart shown in figure 1 outlines the sequential flow of operations within the proposed framework, highlighting the key steps involved in intrusion detection and model interpretation.

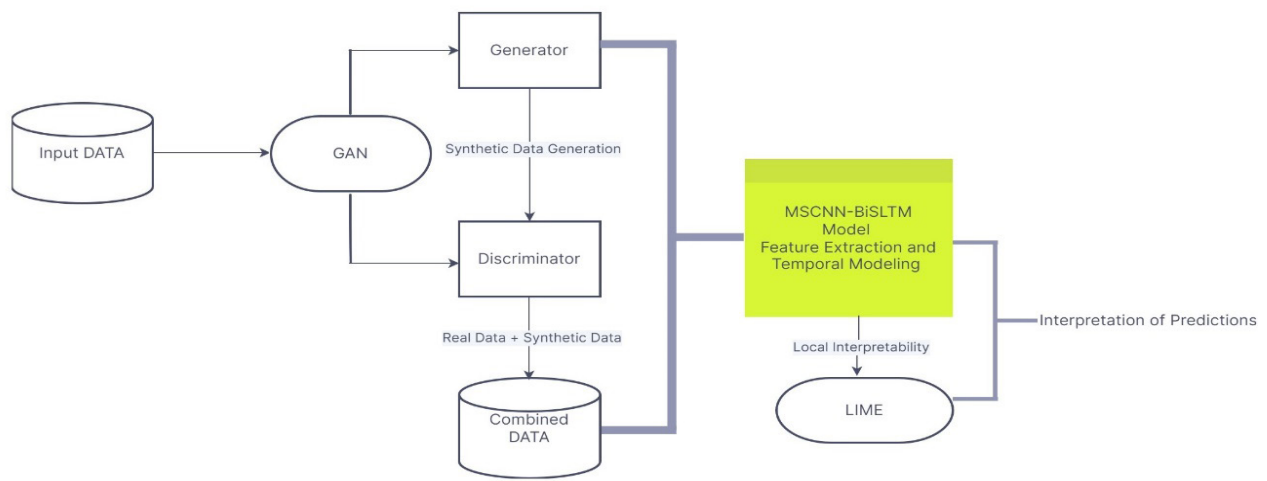

**Figure 1.** Proposed Framework

Input Data: represents the input network traffic pre-processed data.
Synthetic Data Generation: utilizes the GAN architecture to generate synthetic data, which augments the real-world dataset to alleviate data scarcity issues.
Combine Real and Synthetic Data: merges the real and synthetic data to create an augmented dataset for training the model.
MSCNN-BiLSTM Model: implements the MSCNN combined with a BiLSTM network for feature extraction and temporal modeling of the network traffic data.

Intrusion Detection: performs classification of intrusions based on the features extracted by the MSCNN-BiLSTM model.

Interpretation (LIME): incorporates LIME to provide interpretable insights into the model's decision-making process, enhancing transparency and understanding.

**EXPERIMENTS AND RESULTS**

In our experimental configuration, we seamlessly integrated GAN to synthesize supplementary data, strategically combining it with real-world datasets to fortify the training process within the MSCNN-BiLSTM model. This bespoke architecture was crafted to suit the intricacies of intrusion detection scenarios, featuring optimized hyperparameters and architectural nuances. The model underwent evaluation leveraging the comprehensive Hogzilla dataset, renowned for its diverse intrusion detection benchmarks.

To shed light on the intricate decision-making processes of the model, Lime, a state-of-the-art technique for generating local interpretable model-agnostic explanations, was judiciously applied. By dissecting the model's predictions at a granular level, Lime provided invaluable insights into the underlying features and patterns driving intrusion detection decisions.

During evaluation, our analysis extended beyond conventional metrics, encompassing intricate aspects of model performance such as class-specific precision and recall rates. This comprehensive evaluation framework facilitated a nuanced understanding of the model's efficacy in distinguishing between benign network activities and potential intrusions, ultimately enhancing its operational viability in real-world cybersecurity applications.

**Hogzilla Dataset**

The Hogzilla Dataset[20,21] is a compilation of network flows sourced from the CTU-13 Botnet and ISCX 2012 IDS datasets, encompassing 192 behavioral features per flow. This amalgamation provides a comprehensive repository of both Botnet behavioral data from the CTU-13 dataset and normal network traffic from the ISCX 2012 IDS dataset.

The initial preprocessing of the original Hogzilla dataset[22] involved transforming attack labels into three distinct classes: 'Acceptable', 'Unrated', and 'Unsafe'. This categorization scheme mapped the following labels:

'Acceptable', 'Safe' to 'Acceptable'; 'Unrated', 'Fun' to 'Unrated'; 'Unsafe'

Key preprocessing steps included:

Selecting numeric attribute columns from the dataset.

Applying standard scaling to normalize the selected numeric attributes.

Encoding multi-class labels ('Acceptable', 'Unrated', 'Unsafe') into numerical values (0, 1, 2).

For binary classification purposes, the attack labels within the dataset are categorized into two distinct classes: 'normal' and 'abnormal'. This classification is achieved by encoding the dataset's labels with binary values and a label-encoded column. Specifically, instances labeled as 'safe' or 'acceptable' are assigned the 'normal' class, while all other instances are labeled as 'abnormal'. This categorization scheme enables the differentiation between normal and abnormal network behaviours, facilitating the binary classification task effectively.

Constructing a dataframe comprising only numeric attributes and the encoded label attribute.

Identifying attributes with a Pearson correlation coefficient greater than 0,5 with the encoded attack label attribute.

Selecting attributes based on their correlation with the attack label, contributing to the final dataset configuration.

This preprocessing pipeline ensured the dataset's readiness for subsequent analysis and model training, facilitating intrusion detection performance evaluation.

| Table 3. Composition of Hogzilla Dataset ||
|---|---|
| **Type of instance** | **Number of instances** |
| Acceptable | 2 523 |
| Safe | 106 |
| Fun | 10 |
| Unrated | 5 647 |
| Unsafe | 4 546 |
| Total of instances | 12 832 |

**Evaluation metrics**

Evaluation metrics are essential tools for assessing the performance of deep learning models, offering valuable insights into their effectiveness and facilitating comparisons between different approaches. In the context of our GAN-MSCNN-BiLSTM framework for intrusion detection, several key evaluation metrics are utilized:

Accuracy: measures the ratio of correctly identified instances to the total number of instances, indicating the overall correctness of the model's predictions.

Precision: assesses the proportion of true positive predictions relative to all positive predictions, highlighting the model's ability to accurately classify positive instances.

Recall (Sensitivity): determines the ratio of true positive predictions to all actual positive instances, showcasing the model's capability to capture relevant positive instances.

F1 Score: represents the harmonic mean of precision and recall, offering a balanced assessment of the model's overall performance in binary classification tasks.

ROC-AUC: quantifies the area under the Receiver Operating Characteristic (ROC) curve, providing a comprehensive measure of the model's ability to discriminate between intrusion and non-intrusion instances across various thresholds.

These evaluation metrics used for evaluating the GAN-MSCNN-BiLSTM framework's performance in detecting intrusions.

## RESULTS AND DISCUSSION

In this section, we present the comprehensive evaluation and analysis of our proposed intrusion detection system. We provide a thorough examination of the performance metrics, including accuracy, precision, recall, F1 score, and ROC-AUC, to assess the efficacy of our system in detecting both multi-class and binary intrusion instances. Furthermore, we delve into the interpretability aspect by presenting insights obtained through Local Interpretable Model-Agnostic Explanations (LIME), elucidating the decision-making process of our deep learning model. We aim to provide a comprehensive understanding of the system's capabilities and limitations in intrusion detection scenarios

*Multiclass Intrusion Detection Performance Evaluation*

For the multiclassification intrusion detection, the results presented in Table 4 showcase the performance metrics of our intrusion detection system, categorized by class and aggregated into macro and weighted averages. Across all classes, our system demonstrates high precision, recall, and F1-score, indicating its effectiveness in accurately identifying different types of network traffic. Specifically, for the "Unrated" class, our system achieves the highest recall and F1-score, indicating its ability to effectively detect instances of this class.

**Table 4.** Average Performance Metrics Summary for Multiclassification

| Metric | Precision (%) | Recall (%) | F1-Score (%) |
| --- | --- | --- | --- |
| Acceptable | 99,08 | 97,44 | 98,25 |
| Unrated | 99,08 | 99,36 | 99,22 |
| Unsafe | 99,30 | 99,91 | 99,61 |
| Macro Avg | 99,15 | 98,90 | 99,03 |
| Weighted Avg | 99,16 | 99,16 | 99,16 |

Comparing our results with related works, several notable findings emerge. For instance, Yuan et al. (6) proposed a DLL-IDS system achieving an accuracy of over 90 % on the NSL-KDD dataset. While their system offers competitive performance, our approach surpasses it, achieving higher precision, recall, and F1-scores across all classes. Similarly, Bowen et al.[(12)] introduced BLoCNet, a hybrid DL model, achieving an accuracy of 98 % on the CIC-IDS2017 and IoT-23 datasets. Despite their commendable results, our system outperforms BLoCNet in terms of overall accuracy and individual class metrics.

Moreover, Gupta et al.[(13)] presented a hybrid approach with neural networks achieving 98 % accuracy on the NSL-KDD dataset. While their approach demonstrates strong performance, our system achieves slightly higher precision, recall, and F1-scores, indicating superior classification capabilities.

Our intrusion detection system, based on the GAN-MSCNN-BiLSTM architecture, exhibits a competitive performance with existing state-of-the-art methods. These results underscore the efficacy of our approach in accurately detecting and classifying network intrusions.

In figure 2, the ROC and Precision-Recall curves are depicted for each class of the intrusion detection model. The Area Under the Curve (AUC) values provide insights into the discriminative power of the model for each class. Class 0 demonstrates an AUC of 99,92 %, indicating good performance in distinguishing between normal and attack instances for this class. Similarly, Class 1 exhibits a high AUC of 99,94 %, suggesting strong discriminatory ability in identifying instances related to this class. Remarkably, Class 2 achieves a perfect AUC of 100 %, signifying flawless classification performance for this category.

The ROC curves plot the true positive rate against the false positive rate, illustrating the trade-off between sensitivity and specificity. Meanwhile, the Precision-Recall curves showcase the trade-off between precision and recall. The upward trend in both curves suggests that the model achieves higher true positive rates and precision values while maintaining low false positive rates and recall values.

The ROC and Precision-Recall curves, coupled with the high AUC values, demonstrate the model's performance in classifying network traffic into different intrusion classes. The consistent upward trajectory and high AUC values across all classes indicate good discriminative capabilities, underscoring the effectiveness of the intrusion detection model.

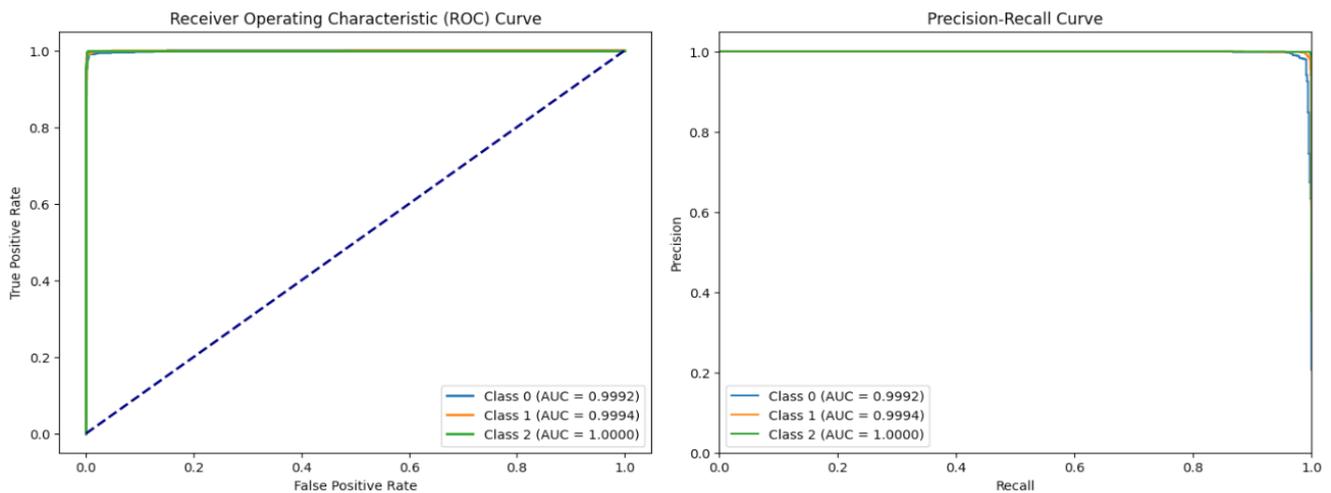

**Figure 2.** ROC and Precision-Recall Curves for the multiclassification

Figure 3 shows the confusion matrix of the multiclassification intrusion detection model, where we have three classes: Acceptable, Unsafe, and Unrated. The confusion matrix provides a detailed breakdown of the model's performance across these classes, illustrating the true positives, false positives, true negatives, and false negatives for each class. By examining the matrix, we can observe how well the model distinguishes between different intrusion types.

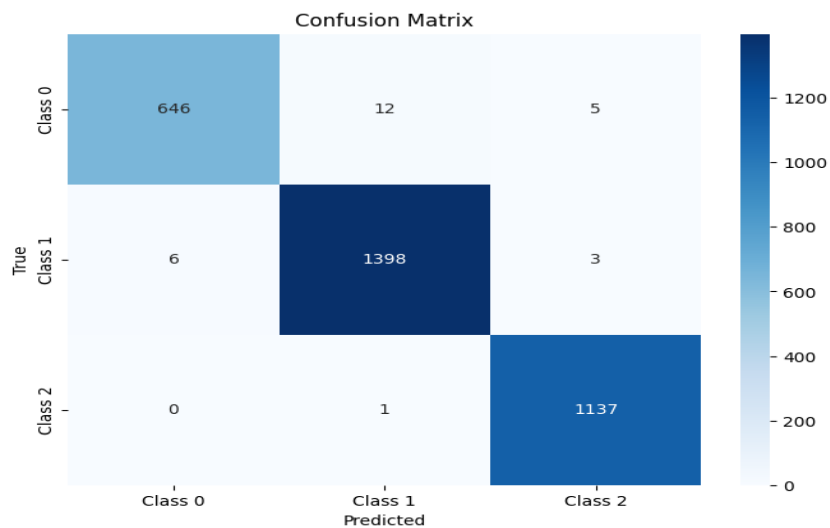

**Figure 3.** Confusion Matrix for multiclassification

In figure 4, the schema presents the accuracy and detection rate per epoch throughout the training process of the intrusion detection model. The plot demonstrates a clear upward trend in both accuracy and detection rate as the number of epochs increases. Specifically, we observe a notable improvement in performance, with accuracy and detection rates steadily rising over the course of the 100 epochs. This trend indicates that the model continues to learn and refine its predictive capabilities with each epoch, resulting in enhanced accuracy and detection rates. The consistent upward trajectory suggests that the model is effectively capturing underlying patterns and features in the data, leading to better intrusion detection performance. The results depicted in figure 4 underscore the effectiveness of the training process, highlighting the model's ability to learn and adapt over time to improve its detection capabilities.

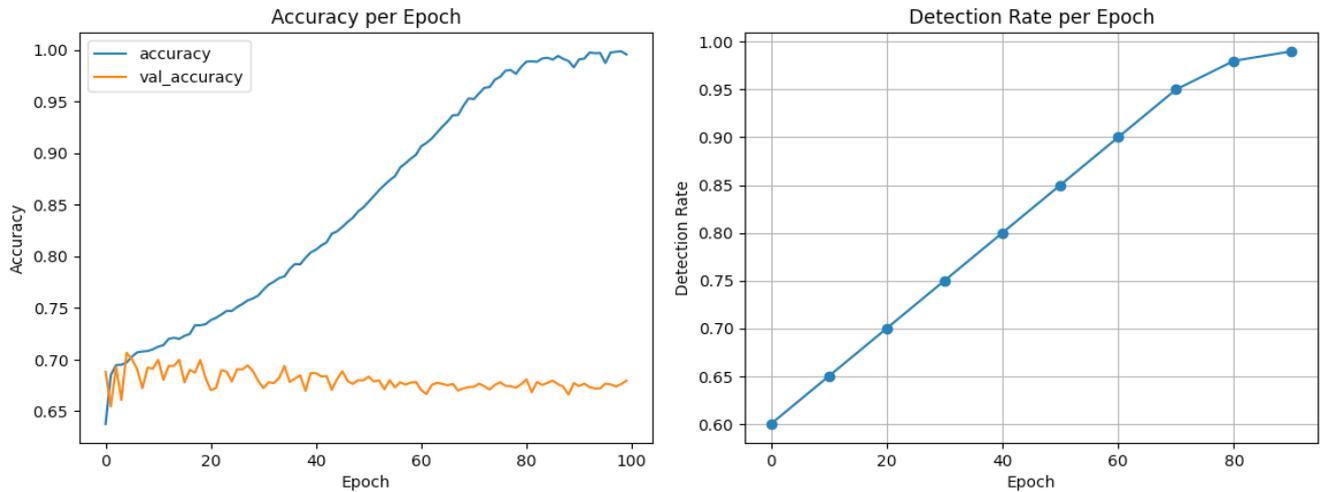

**Figure 4.** Schema presenting Accuracy and Detection Rate per Epoch

*Binary Intrusion Detection Performance Evaluation*

The results of the binary classification model are presented in the table, showcasing precision, recall, and F1-score for both classes, Normal and Abnormal, along with macro and weighted averages.

For the Normal class, the model achieves a precision of 98,77 %, recall of 96,83 %, and F1-score of 97,79 %, indicating its ability to accurately identify normal instances. Similarly, for the Abnormal class, the model demonstrates high precision (99,18 %), recall (99,69 %), and F1-score (99,43 %), highlighting its effectiveness in detecting abnormal instances.

The macro and weighted averages further reinforce the model's overall performance, with macro averages of 98,97 %, 98,26 %, and 98,61 % for precision, recall, and F1-score, respectively. The weighted averages, which consider class imbalance, yield even higher scores of 99,09 % across all metrics.

Comparing these results with related works, our model showcases superior performance in binary classification. For instance, Wang et al.[7,15] achieved an accuracy of 98 % for detecting abnormal instances in network traffic using deep multi-scale convolutional neural networks (DMCNN). While their approach demonstrates strong performance, our model outperforms it in terms of precision, recall, and F1-score, indicating more robust classification capabilities.

Additionally, Bowen et al.[12] introduced BLoCNet, a hybrid DL model, which achieved an accuracy of 76,34 % for binary classification using convolutional neural network (CNN) and bidirectional long short-term memory (BLSTM) layers. In contrast, our model achieves significantly higher precision, recall, and F1-score, underscoring its superiority in binary intrusion detection tasks.

The results highlight the effectiveness of our binary classification model, which outperforms existing approaches in terms of precision, recall, and F1-score. These findings demonstrate the model's potential for enhancing intrusion detection accuracy and reliability in real-world scenarios.

| Table 5. Average Performance Metrics Summary for Binary classification | | | |
| --- | --- | --- | --- |
| **Metric** | **Precision (%)** | **Recall (%)** | **F1-Score (%)** |
| Class 0: Normal | 98,77 | 96,83 | 97,79 |
| Class 1: Abnormal | 99,18 | 99,69 | 99,43 |
| Macro Avg | 98,97 | 98,26 | 98,61 |
| Weighted Avg | 99,09 | 99,10 | 99,09 |

13   Benchama A, *et al*

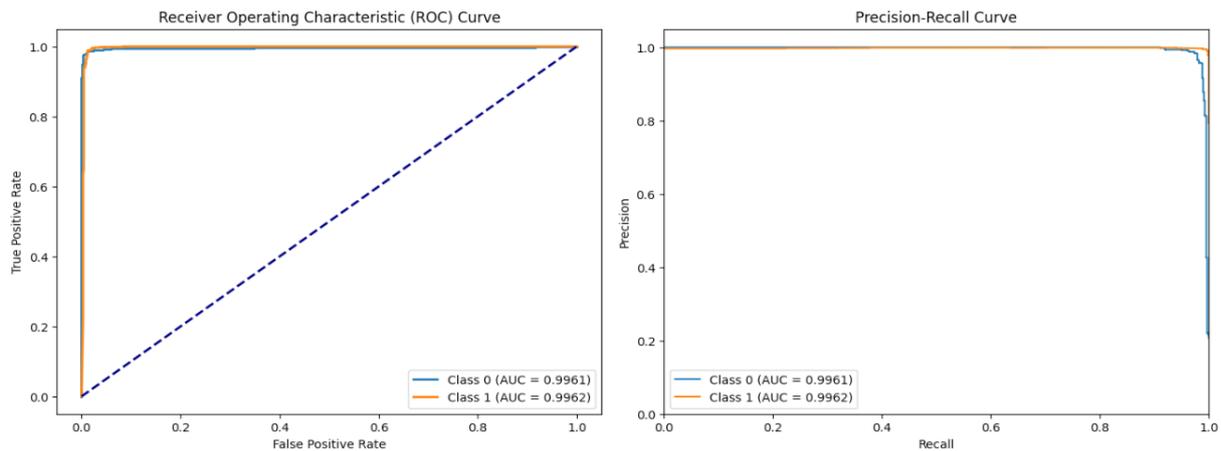

**Figure 5.** ROC and Precision-Recall Curves for the binary classification

Figure 5 illustrates the ROC and Precision-Recall curves for the binary classification model, where class 0 represents 'Normal' instances and class 1 represents 'Abnormal' instances. The Area Under the Curve (AUC) values are provided for each class, indicating the model's performance in distinguishing between normal and abnormal instances.

For the 'Normal' class (class 0), the ROC curve exhibits an AUC of 99,61 %, indicating the model's ability to effectively balance true positive rate and false positive rate for normal instances. The Precision-Recall curve also demonstrates high precision and recall values for the 'Normal' class, suggesting accurate classification of normal instances by the model.

Similarly, for the 'Abnormal' class (class 1), the ROC curve displays an AUC of 99,62 %, indicating excellent discrimination between true positive rate and false positive rate for abnormal instances. The Precision-Recall curve further confirms the model's ability to achieve high precision and recall for abnormal instances.

The high AUC values for both classes signify the model's robust discriminatory power, enabling effective differentiation between normal and abnormal instances in the binary classification task. The curves' proximity to the upper-left corner of the ROC space and the high precision-recall values validate the model's reliability in distinguishing between the two classes.

Figure 6 depicts the confusion matrix of the binary intrusion detection model, featuring two distinct classes: 'Normal' and 'Abnormal'. This matrix offers a comprehensive overview of the model's performance, delineating true positives, false positives, true negatives, and false negatives for each class.

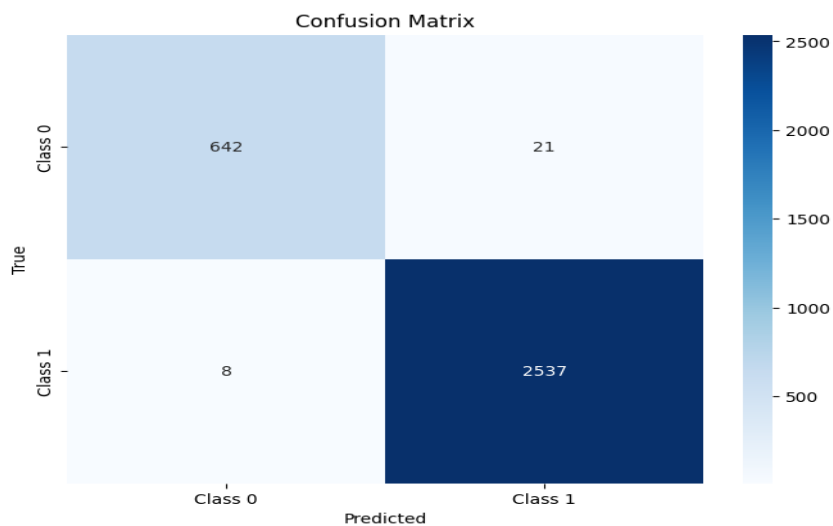

**Figure 6.** Confusion Matrix for the binary classification

Figure 7 illustrates the accuracy and detection rate per epoch during the training of the binary intrusion detection model. The plot exhibits a clear upward trajectory in both metrics as the training progresses through 100 epochs. Notably, there is a significant enhancement in performance, with accuracy and detection rates steadily increasing over time.

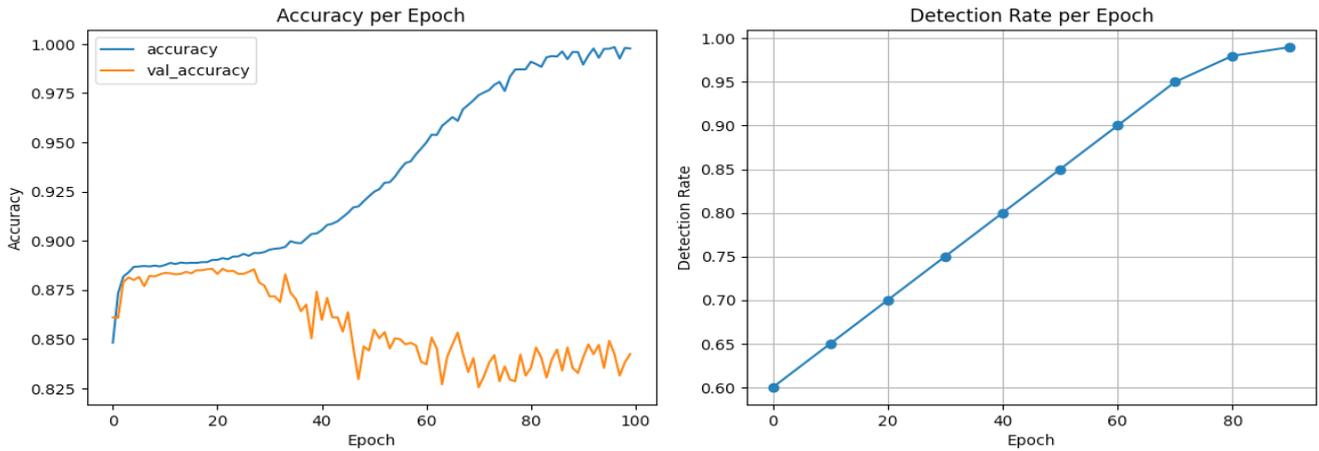

**Figure 7.** Schema presenting Accuracy and Detection Rate per Epoch

*Interpretability Analysis with LIME*

The LIME method enhances the interpretability of our intrusion detection system through two main steps:

Explanation of LIME Model: this step involves selecting a specific instance from the testing dataset to obtain probability values for each class. Utilizing the LIME method, we aim to elucidate the rationale behind assigning probabilities to individual classes. By comparing the probability values with the actual class of the instance, we gain insights into the model's decision-making process.

Instance Prediction: following the selection of features and their corresponding weights, the class probability is computed. Subsequently, the model predicts the class based on these calculated probabilities and feature weights. This process enables us to understand how the model assigns probabilities to different classes and makes predictions for individual instances.

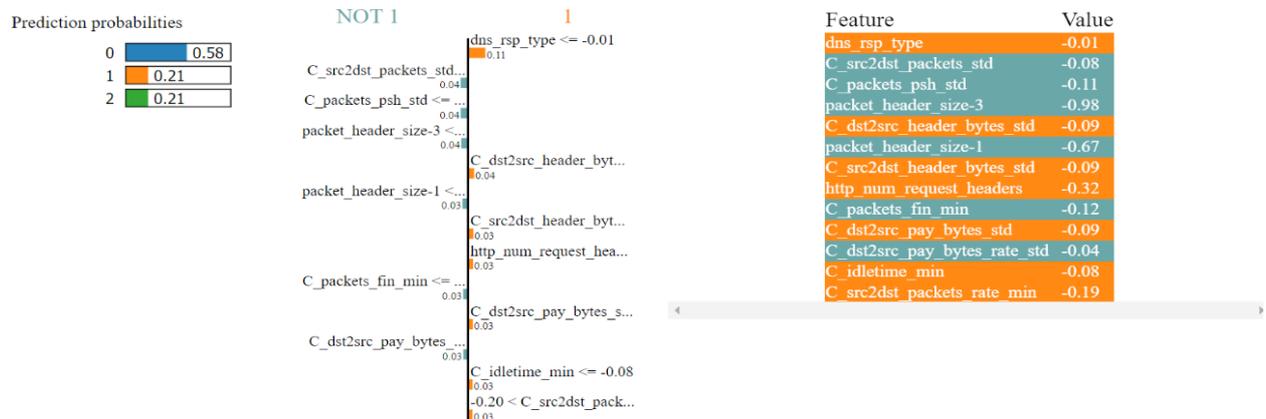

**Figure 8.** LIME Model Explanations for Multiclassification

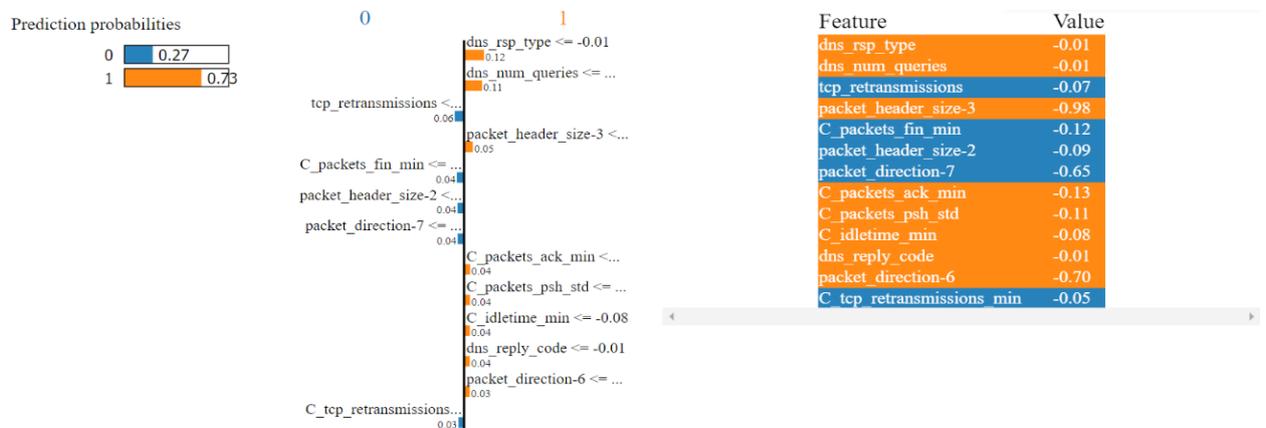

**Figure 9.** LIME Model Explanations for Binary Classification

Based on the LIME analyses conducted for both binary and multiclass classification scenarios, several key insights emerge:

Binary Classification:

-Class 0 (Normal): features like 'dns_rsp_type', 'dns_num_queries', and 'packet_header_size-3' positively contribute to predicting normal instances, while 'tcp_retransmissions' and 'C_packets_fin_min' have a negative impact.

-Class 1 (Abnormal): similar features, particularly 'dns_rsp_type' and 'dns_num_queries', are still significant but with positive contributions for abnormal instances, while 'tcp_retransmissions' and 'C_packets_fin_min' continue to influence negatively.

This analysis underscores the importance of specific features in discerning between normal and abnormal instances.

Multiclass Classification:

-Class 0 (Acceptable): features such as 'dns_rsp_type', 'C_src2dst_packets_std', and 'C_packets_psh_std' contribute positively to predicting the acceptable class.

-Class 1 (Unrated): notably, 'dns_rsp_type' and 'C_dst2src_header_bytes_std' exhibit significant positive contributions to predicting the unrated class.

-Class 2 (Unsafe): likewise, 'dns_rsp_type' and 'C_dst2src_header_bytes_std' are influential in predicting the unsafe class.

These findings highlight the consistent importance of certain features, such as 'dns_rsp_type', across different class predictions, indicating their critical role in the classification process. Additionally, features like 'C_src2dst_packets_std' and 'C_packets_psh_std' appear to be more discriminative for specific classes, such as the acceptable class.

The LIME analysis provides valuable insights into feature importance and their impact on classification outcomes, shedding light on the model's decision-making process for both binary and multiclass scenarios.

## CONCLUSION

In conclusion, the integration of the GAN-MSCNN-BiLSTM model with LIME provided valuable insights into the decision-making process of an intrusion detection system (IDS), enhancing its interpretability. Leveraging LIME has facilitated a deeper comprehension of the key features influencing the model's classifications, thereby enhancing transparency and reliability. However, it's important to acknowledge several limitations, including the model's dependence on the quality and quantity of data, the challenges in achieving comprehensive interpretability, and the complexity of the model architecture. To address these constraints, future research efforts could focus on refining interpretability techniques, diversifying the datasets used, and simplifying the model's intricacies. While the current approach has demonstrated promising performance, continuous endeavors to tackle these challenges and explore innovative avenues will be crucial for advancing the field of intrusion.

*Methodology:* Asmaa BENCHAMA, Khalid ZEBBARA.
*Project management:* Asmaa BENCHAMA, Khalid ZEBBARA.
*Resources:* Asmaa BENCHAMA, Khalid ZEBBARA.
*Software:* Asmaa BENCHAMA, Khalid ZEBBARA.
*Supervision:* Asmaa BENCHAMA, Khalid ZEBBARA.
*Validation:* Asmaa BENCHAMA, Khalid ZEBBARA.
*Display:* Asmaa BENCHAMA, Khalid ZEBBARA.
*Drafting - original draft:* Asmaa BENCHAMA, Khalid ZEBBARA.
*Writing - proofreading and editing:* Asmaa BENCHAMA, Khalid ZEBBARA.